\documentclass[12pt,a4paper]{article}
\usepackage{a4wide}
\usepackage{amsmath,amssymb,amsfonts,amsxtra,simpler-wick} 
\usepackage{cite}

\allowdisplaybreaks

\begin{document}
\thispagestyle{empty}

\vbox{\noindent November 2022 \hfill IPMU22-0053}

\vskip 3.0cm

\begin{center}{\Large \bf Schwarzschild-type black holes in Starobinsky-Bel-Robinson gravity}\end{center}  

\vspace*{1cm}

\centerline{Ruben Campos Delgado$^{a,}$\footnote{E-mail: ruben.camposdelgado@gmail.com} and Sergei V. Ketov$^{b,c,d,}$\footnote{E-mail: ketov@tmu.ac.jp} }

\vspace{1cm}

\begin{center}{\it
$^a$Bethe Center for Theoretical Physics,\\
Physikalisches Institut der Universit\"at Bonn,\\
Nussallee 12, 53115 Bonn, Germany\\
$^b$Department of Physics, Tokyo Metropolitan University, \\
1-1 Minami-ohsawa, Hachioji-shi, Tokyo 192-0397, Japan\\
$^c$Interdisciplinary Research Laboratory, \\Tomsk State University, 
36 Lenin Avenue, Tomsk 634050, Russia\\
$^c$Kavli Institute for the Physics and Mathematics of the Universe,\\
The University of Tokyo Institutes for Advanced Study, Kashiwa 277-8583, Japan}
\end{center}

\vspace*{1cm}

\centerline{\bf Abstract}
\vskip .3cm

We study physical properties of a Schwarzschild-type black hole in the framework of the recently proposed Starobinsky-Bel-Robinson (SBR) modified theory of gravity, working perturbatively in the new coupling constant. In particular, we compute the temperature, entropy, pressure and lifetime of a Schwarzschild-type black hole.


\vskip .3cm

\newpage
\section{Introduction}\label{sec:intro}
The Starobinsky model of modified $(R+R^2)$ gravity \cite{Starobinsky:1980te} is known as an excellent model of cosmological inflation, in very good agreement with WMAP/Planck/BICEP/KECK precision measurements of the cosmic microwave background radiation \cite{BICEP:2021xfz}.~\footnote{ See e.g., Ref.~\cite{Ketov:2022qwj} for a recent review.} The Starobinsky model has no free parameters because its only parameter given by the inflaton mass $m$ is determined by the COBE/WMAP normalization as $m\sim 10^{-5}M_{\rm Pl}$ in terms of the reduced Planck mass $M_{\rm Pl}=1/\sqrt{8\pi G_N}$. The ultra-violet (UV) cutoff of the Starobinsky model is given by $M_{\rm Pl}$ also, while the Starobinsky inflation belongs to the class of the single-large-field inflationary models that are UV-sensitive to quantum gravity corrections. It is therefore of importance to identify those corrections and study their applications in the high-curvature regimes relevant to the early universe and black holes.

A perturbative extension of the Starobinsky $(R+R^2)$ gravity in $D=4$ dimensions by the terms {\it quartic} in the curvature was proposed in Ref.~\cite{Ketov:2022lhx}. This extension was inspired by the leading quantum gravity
correction obtained by dimensional reduction of M-theory in $D=11$ dimensions. The extra terms in the $D=4$ gravitational low-energy effective action have the peculiar structure given by square of the Bel-Robinson tensor \cite{Bel:1959uwe, Robinson:1997}, so the new modified gravity was dubbed the
 Starobinsky-Bel-Robinson (SBR) gravity in Ref.~~\cite{Ketov:2022lhx}. The SBR action reads
\begin{equation}\label{eq:sbr_action}
S_{\rm SBR}[g_{\mu\nu}]=\frac{M^2_{\text{Pl}}}{2}\int d^4x \sqrt{-g}\, \left[R+\frac{1}{6m^2}R^2-\frac{\beta}{32M^6_{\text{Pl}}}\left(\mathcal{P}^2-\mathcal{G}^2\right)\right],
\end{equation}
where $\beta>0$ is the new dimensionless coupling constant whose (unknown) value is supposed to be determined by compactification of M-theory, $\mathcal{G}$ and $\mathcal{P}$ are the Euler and Pontryagin topological densities in $D=4$, respectively, coming from the BR-tensor squared,
\begin{equation}
    \mathcal{G}=R^2-4R_{\mu\nu}R^{\mu\nu}+R_{\mu\nu\rho\sigma}R^{\mu\nu\rho\sigma}
\end{equation}
and
\begin{equation}
    \mathcal{P}=\frac{1}{2}\sqrt{-g}\epsilon_{\mu\nu\rho\sigma}{R^{\rho\sigma}}_{\alpha\beta}R^{\mu\nu\alpha\beta}.
\end{equation}
The SBR action (\ref{eq:sbr_action}) can be rewritten in the equivalent form
\begin{equation}\label{Slfields}
S_{\rm SBR}[g_{\mu\nu},\phi,\chi,\xi]=\frac{M^2_{\rm Pl}}{2}S_{R}-\frac{\beta}{32M_{\rm Pl}^4}\left(S_{\cal G}+S_{\cal P}\right),
\end{equation}
where we have introduced the scalar fields $\phi$, $ \chi$ and $\xi$, together with
\begin{eqnarray}
S_{R}[g_{\mu\nu},\phi]&=&\int d^4x\sqrt{-g}\left[R\left(1+\frac{\phi}{3 m^2}\right)-\frac{\phi^2}{6m^2}\right], \nonumber \\
S_{\cal{G}}[g_{\mu\nu},\chi]&=&\int d^4x\sqrt{-g}\left(\frac{\chi^2}{2}-{\cal G}\chi\right),\nonumber \\
S_{\cal P}[g_{\mu\nu},\xi]&=&\int d^4x\sqrt{-g}\left(\xi \mathcal{P}-\frac{\xi^2}{2}\right). 
\end{eqnarray}
The scalar field $\phi$ is related to Starobinsky's scalaron, the scalar field $\chi$ is related to string dilaton and the pseudo-scalar field $\xi$ is related to string axion, in the absence of their kinetic terms. Despite the presence of the higher derivatives beyond the second order, the non-perturbative Starobinsky $(R+\frac{1}{6m^2}R^2)$ gravity is known to be ghost-free for $R>-3m^2$. The dilaton-axion couplings to the spacetime topological densities are also ghost-free in string theory. Unlike Ref.~\cite{Crisostomi:2017ugk}, we treat the Chern-Simons-type ${\cal G}$- and $\mathcal{P}$-couplings as the perturbative ones in the possible low-energy part of the effective action of a more fundamental theory, so that  they do not introduce any new degrees of freedom beyond those present in the Starobinsky gravity.

In this letter we study how the physical properties of a Schwarzschild-type black hole are modified by the presence of the 
$\beta$-dependent terms in Eq.~(\ref{eq:sbr_action}). Since the value of the $\beta$-parameter is assumed to be small,
$\beta\ll 1$,  we work perturbatively in the first order with respect to $\beta$. We also assume that the spacetime curvature is much less that $M^2_{\rm Pl}$. These conditions guarantee the absence of ghosts and consistency of our approach, see also 
Ref.~\cite{Ketov:2022zhp} for additional checks and physical applications.

Our paper is organized as follows. In Section \ref{sec:sec2} we compute the $\beta$-dependent correction to the Schwarzschild metric. This correction shifts the position of the event horizon. Having the displaced event horizon leads to the corresponding change in the black hole entropy obtained by integrating over the horizon area. In Section  \ref{sec:sec3}  we derive the $\beta$-corrections to the temperature, pressure and lifetime of a Schwarzschild-type  black hole. Section \ref{sec:sec4} is our conclusion. We use natural units with $\hbar=c=k_B=1$.
\section{$\beta$-correction to the Schwarzschild metric}\label{sec:sec2}
 In this Section we derive and solve the equations of motion obtained by varying the action \eqref{eq:sbr_action} with respect to the metric. Our task is simplified by the fact that the Pontryagin density $\mathcal{P}$ vanishes for a spherically symmetric metric.  The equations of motion are
\begin{equation}\label{eq:structure_of_equations}
    J_{\mu\nu}:=G_{\mu\nu}+\frac{1}{6m^2}H_{\mu\nu}-\frac{\beta}{32M^6_{\text{Pl}}} K_{\mu\nu}=0~~,
\end{equation}
where $G_{\mu\nu}=R_{\mu\nu}-1/2 g_{\mu\nu}R$ is the Einstein tensor,
\begin{equation}
    H_{\mu\nu}=2R_{\mu\nu}R - \frac{1}{2}g_{\mu\nu}R^2-2\nabla_{\mu}\nabla_{\nu}R+2g_{\mu\nu}\Box{R}
\end{equation}
and
\begin{equation}
\begin{gathered}
    K_{\mu\nu}=\frac{1}{2}g_{\mu\nu}\mathcal{G}^2-2\Big[2\mathcal{G}RR_{\mu\nu}-4\mathcal{G}{R_{\mu}}^{\rho}R_{\nu\rho}\\+2\mathcal{G}{R_{\mu}}^{\rho\sigma\lambda}R_{\nu\rho\sigma\lambda}
    +4\mathcal{G}R^{\rho\sigma}R_{\mu\rho\sigma\nu}+2g_{\mu\nu}R\Box{\mathcal{G}}-2R\nabla_{\mu}\nabla_{\nu}\mathcal{G}\\-4R_{\mu\nu}\Box\mathcal{G}
    +4\left(R_{\mu\rho}\nabla^{\rho}\nabla_{\nu}\mathcal{G}+R_{\nu\rho}\nabla^{\rho}\nabla_{\mu}\mathcal{G}\right)\\-4g_{\mu\nu}R_{\rho\sigma}\nabla^{\sigma}\nabla^{\rho}\mathcal{G}+4R_{\mu\rho\nu\sigma}\nabla^{\sigma}\nabla^{\rho}\mathcal{G}\Big].
\end{gathered}
\end{equation}
We search for a solution in the form
\begin{equation} \label{eq:metric}
    ds^2=-a(r)dt^2+\frac{1}{a(r)}dr^2+r^2d\Omega^2_2,
\end{equation}
where
\begin{equation}
    d\Omega^2_2=d\theta^2+\sin^2\theta d\phi^2~,
\end{equation}
and with the function $a(r)$ beginning with the Schwarzschild solution
\begin{equation}\label{eq:ansatz}
    a(r)=1-\frac{2G_NM}{r}-\beta f(r),
\end{equation}
in the presence of a perturbation $-\beta f(r)$. We assume that $\beta \ll 1$ and linearize our equations with respect to
$\beta$.

Because of spherical symmetry, there are only four independent equations to consider. Plugging \eqref{eq:ansatz} into \eqref{eq:structure_of_equations} we find, at first order in $\beta$, 
\begin{equation}\label{eq:jtt}
\begin{gathered}
    J_{tt}=\frac{r-2G_NM}{r^{13}}\bigg[\frac{36G^3_NM^3}{M^6_{\text{Pl}}}\left(67G_NM-32r\right)\\
    +r^7\left(r^3+\frac{4G_NM}{m^2}-\frac{4r}{3m^2}\right)f(r)
    +r^8\left(r^3-\frac{2G_NM}{m^2}+\frac{4r}{3m^2}\right)f'(r)\\
    -\frac{2r^{10}}{3m^2}f''(r)+\frac{r^{10}}{3m^2}\left(11G_NM-6r\right)f^{(3)}(r)\\
    +\frac{r^{11}}{3m^2}\left(2G_NM-r\right)f^{(4)}(r)\bigg]\beta + \mathcal{O}\left(\beta^2\right)=0,
\end{gathered}
\end{equation}
\begin{equation}
\begin{gathered}\label{eq:jrr}
    J_{rr}= \frac{1}{(2G_NM-r)r^{11}}\bigg[\frac{36G^3_NM^3}{M^6_{\text{Pl}}}\left(11G_NM-4r\right) \\
    + r^7\left(r^3-\frac{4G_NM}{m^2}+\frac{8r}{3m^2}\right)f(r)
    +r^8\left(r^3-\frac{2G_NM}{m^2}+\frac{4r}{3m^2}\right)f'(r)\\
    +\frac{4r^9}{3m^2}\left(3G_NM-2r\right)f''(r)
    +\frac{r^{10}}{3m^2}\left(3G_NM -2r\right)f^{(3)}(r)\bigg]\beta + \mathcal{O}\left(\beta^2\right)=0,
\end{gathered}
\end{equation}
\begin{equation}\label{eq:jthetatheta}
\begin{gathered}
    J_{\theta\theta}=\frac{1}{2r^{10}}\bigg[\frac{72G^3_NM^3}{M^6_{\text{Pl}}}\left(41G_NM-18r\right)\\+\frac{16r^7}{3m^2}\left(r-3G_NM\right)f(r)-2r^9\left(r^2+\frac{2}{3m^2}\right)f'(r)\\
    -r^9\left(r^3+\frac{4}{3m^2}(r-6G_NM)\right)f''(r)
    +\frac{2r^{10}}{3m^2}\left(5r-8G_NM\right)f^{(3)}(r)\\
    +\frac{2r^{11}}{3m^2}\left(r-2G_NM\right)f^{(4)}(r)\bigg]\beta+\mathcal{O}\left(\beta^2\right)=0,
\end{gathered}
\end{equation}
\begin{equation}
\begin{gathered}\label{eq:jphiphi}
    J_{\phi\phi}=\frac{\sin^2\theta}{2r^{10}}\bigg[\frac{72G^3_NM^3}{M^6_{\text{Pl}}}\left(41G_NM-18r\right)\\+\frac{16r^7}{3m^2}\left(r-3G_NM\right)f(r)-2r^9\left(r^2+\frac{2}{3m^2}\right)f'(r)\\
    -r^9\left(r^3+\frac{4}{3m^2}(r-6G_NM)\right)f''(r)
    +\frac{2r^{10}}{3m^2}\left(5r-8G_NM\right)f^{(3)}(r)\\
    +\frac{2r^{11}}{3m^2}\left(r-2G_NM\right)f^{(4)}(r)\bigg]\beta+\mathcal{O}\left(\beta^2\right)=0\bigg].
\end{gathered}
\end{equation}

This system of differential equations can be reduced to a single equation after eliminating the 3rd and 4th order derivatives. Indeed, solving $j_{\theta\theta}$ or $j_{\phi\phi}$ for $f^{(4)}(r)$, $j_{rr}$ for $f^{(3)}(r)$ and plugging the resulting expressions into $j_{tt}$, we obtain the equation
\begin{equation}
\begin{gathered}
    \frac{72G^3_NM^3}{M^6_{\text{Pl}}}\left(291G^2_NM^2-343G_NMr+96r^2\right)\\
    -2r^{11}f(r)+2r^{11}\left(r-3G_NM\right)f'(r)
    +r^{12}\left(2r-3G_NM\right)f''(r)=0~,
\end{gathered}
\end{equation}
whose solution is
\begin{equation}\label{eq:solution_f(r)}
\begin{gathered}
    f(r)=\frac{4G^3_NM^3}{5M^6_{\text{Pl}}r^{10}}\left(97G_NM-54r\right)
    =\frac{2048\pi^3 G^6_NM^3}{5r^{10}}\left(97G_NM-54r\right).
\end{gathered}
\end{equation}
As a result, we find the function $a(r)$ in the form
\begin{equation} \label{eq:solution_a(r)}
    a(r)=1-\frac{r_S}{r} +\beta \frac{128\pi^3}{5}\left(\frac{G_Nr_S}{r^3}\right)^3 \left( 108 - 97 \frac{r_S}{r}\right)~,
\end{equation}
where we have introduced the (unperturbed) Schwarzschild radius $r_S=2G_NM$.
\section{$\beta$-corrections to the physical properties of a \\ Schwarzschild-type black hole}\label{sec:sec3}
Let us now study how the solution \eqref{eq:solution_a(r)} affects the physical properties of a Schwarzschild-type black hole, namely, temperature, entropy, pressure and lifetime. The horizon radius is determined by a solution of $a(r)=0$. At first order in $\beta$ we find
\begin{equation} \label{eq:horizon}
    r_H=2G_NM -\beta\frac{44\pi^3}{5G^2_NM^5}+\mathcal{O}\left(\beta^2\right).
\end{equation}
Therefore, the Hawking temperature is given by
\begin{equation}\label{eq:temperature}
    T_H=\frac{1}{4\pi}\frac{da(r)}{dr}\bigg\rvert_{r=r_H}=\frac{1}{8\pi G_N M}+\beta\frac{\pi^2}{G^4_NM^7}+\mathcal{O}\left(\beta^2\right).
\end{equation}
According to the Wald formula \cite{Wald:1993nt}, the entropy is
\begin{equation}
    S=-2\pi \int_{r=r_H}d\Sigma \,\epsilon_{\mu\nu}\epsilon_{\rho\sigma}\frac{\partial \mathcal{L}}{\partial R_{\mu\nu\rho\sigma}},
\end{equation}
where $d\Sigma=r^2\sin\theta d\theta d\phi$, $\epsilon_{\mu\nu}$ is the Levi-Civita tensor in two dimensions and $\mathcal{L}$ is the Lagrangian density of the theory. In our case we have
\begin{equation}
    \mathcal{L}=\frac{1}{16\pi G_N}\left(R+\frac{1}{6m^2}R^2+16\pi^3 G^3_N\beta\mathcal{G}^2\right).
\end{equation}
The antisymmetry of $\epsilon_{\mu\nu}$ reduces the expression of the entropy to
\begin{equation}
    S=-8\pi A \frac{\partial \mathcal{L}}{\partial R_{rtrt}}\bigg\rvert_{r=r_H},
\end{equation}
where $A=4\pi r^2_H$ is the area enclosed by the event horizon. The partial derivative of the Lagrangian with respect to the Riemann tensor is computed by using the formulae \cite{Delgado:2022pcc}
\begin{equation}
\begin{gathered}
    \frac{\partial R}{\partial R_{\mu\nu\rho\sigma}}=\frac{1}{2}\left(g^{\mu\rho}g^{\nu\sigma}-g^{\mu\sigma}g^{\nu\rho}\right),\\
     \frac{\partial R_{\alpha\beta}R^{\alpha\beta}}{\partial R_{\mu\nu\rho\sigma}}=g^{\mu\rho}R^{\nu\sigma}-g^{\nu\rho}R^{\mu\sigma},\\
      \frac{\partial R_{\alpha\beta\gamma\delta}R^{\alpha\beta\gamma\delta}}{\partial R_{\mu\nu\rho\sigma}}=2R^{\mu\nu\rho\sigma}.
\end{gathered}
\end{equation}
We find
\begin{equation}
\begin{gathered}
     \frac{\partial\mathcal{L}}{\partial R_{rtrt}}=\frac{1}{16\pi G_N}\bigg\{-\frac{1}{2}-\frac{R}{6m^2}\\
     +16\pi^3G^3_N\beta\Big[32g^{rr}R^{tt}R_{\alpha\beta}R^{\alpha\beta}-8\left(g^{rr}R^{tt}R^2-R_{\alpha\beta}R^{\alpha\beta}R\right)\\
     -2R^3-8\left(g^{rr}R^{tt}R_{\alpha\beta\gamma\delta}R^{\alpha\beta\gamma\delta}+2R^{rtrt}R_{\alpha\beta}R^{\alpha\beta}\right)\\
     +2\left(2R^{rtrt}R^2-RR_{\alpha\beta\gamma\delta}R^{\alpha\beta\gamma\delta}\right)+4R^{rtrt}R_{\alpha\beta\gamma\delta}R^{\alpha\beta\gamma\delta}\Big]\bigg\}.
\end{gathered}
\end{equation}
The final expression for the entropy is given by
\begin{equation} \label{eq:entropy}
    S=\frac{A}{4G_N}+\beta\left(\frac{304\pi^4}{5G^2_NM^4}+\frac{624\pi^4}{5m^2G_N^4M^6}\right)+\mathcal{O}\left(\beta^2\right).
\end{equation}
The classical relation $TdS=dM$ receives a correction also. We find
\begin{equation}
    T  dS=dM+\beta\left(\frac{8\pi^3}{G^3_NM^6}-\frac{152\pi^3}{5G^3_NM^6}-\frac{468\pi^3}{5m^2G^5_N M^8}\right)dM.
\end{equation}
The physical interpretation of this result is that the Bel-Robinson (quartic) term generates a pressure for the black hole. The first law of thermodynamics is then given by
\begin{equation}
    TdS-PdV
    =dM+\beta\left(\frac{8\pi^3}{G^3_NM^6}-\frac{152\pi^3}{5G^3_NM^6}-\frac{468\pi^3}{5m^2G^5_N M^8}\right)dM,
\end{equation}
where $P$ is the pressure of the black hole and $V$ is its volume, $V=4/3\pi r^3_H$. We can then identify $TdS=dM$ and obtain
\begin{equation} \label{eq:pressure} 
    P=\beta\left(\frac{7\pi^2}{10G^6_NM^8}+\frac{117\pi^2}{40m^2G^8_NM^{10}}\right)+\mathcal{O}\left(\beta^2\right).
\end{equation}
A non-vanishing pressure for a Schwarzschild-type black hole (it has no pressure classically) is the common feature of quantum gravity \cite{Calmet:2021lny}.

The last physical quantity we want to consider is the mass loss rate due to the Hawking radiation. 
If the black hole mass is sufficiently large, then the temperature is low. Accordingly, all the mass lost in the Hawking process is due to the emission of massless particles. The mass loss rate  is given by the Stefan-Boltzmann law \cite{Hiscock:1990ex}
\begin{equation}\label{eq:boltzmann}
    \frac{dM}{dt}=-\xi T^4\left(\frac{21}{8}\sigma_{\nu}+\sigma_{\gamma}+\sigma_g\right),
\end{equation}
where $\xi=\pi^2/60$ is the Stefan-Boltzmann constant and $\sigma_{\nu}$, $\sigma_{\gamma}$, $\sigma_g$ are the thermally averaged cross sections of the black hole for neutrinos, photons and gravitons, respectively. We neglect the neutrino masses in what follows, and introduce the geometrical optics cross section $\sigma_0$ for the Schwarzschild black hole. Defining also
\begin{equation}
    \rho=\left(\frac{21}{8}\sigma_{\nu}+\sigma_{\gamma}+\sigma_g\right)\sigma^{-1}_0,
\end{equation}
we rewrite \eqref{eq:boltzmann} as follows:
\begin{equation}
    \frac{dM}{dt}=-\xi T^4 \rho \sigma_0~.
\end{equation}
The cross sections for neutrinos, photons and gravitons can be estimated as \cite{Hiscock:1990ex}
\begin{equation}
    \sigma_\nu\approx 0.66852 \sigma_0, \hspace{2 mm} \sigma_{\gamma}\approx 0.24044\sigma_0, \hspace{2mm} \sigma_g\approx 0.02748 \sigma_0,
\end{equation}
while most of the power is emitted in neutrinos. The value of $\rho$ is then $\rho\approx 2.0228$.

Classically, we have $\sigma_0=27\pi G^2_NM^2$. We now derive the $\beta$-dependent correction to the classical result. The emitted particles move along null geodesics governed by the equation
\begin{equation}
    \left(\frac{dr}{d\lambda}\right)^2=E^2-a(r)\frac{J^2}{r^2}~~,
\end{equation}
where $\lambda$ is the affine parameter and $E$, $J$ are the energy and angular momentum, respectively. For the emitted particles to reach the infinity rather than falling back into the black hole horizon, one has to require 
$(dr/d\lambda)^2\geq 0$, i.e.
\begin{equation}
    \frac{1}{l^2}\equiv\frac{E^2}{J^2}\geq \frac{a(r)}{r^2}
\end{equation}
for any $r\geq r_{H}$. The inequality is saturated by the maximal value of $a(r)/r^2$. The solution of
$\partial_r \left(a(r)/r^2\right)=0$ corresponds to the critical, unstable  orbit $r_c$, and the impact factor can be computed as $l_c=r_c/\sqrt{a(r_c)}$ \cite{Xu:2019wak, Xu:2019krv}.
The geometrical optics cross section of the black hole is then given by
\begin{equation}\label{eq:cross_section}
    \sigma_0=\pi l^2_c=\pi\frac{r^2_c}{a(r_c)}~.
\end{equation}
With the solution \eqref{eq:solution_f(r)} we find
\begin{equation}
    \sigma_0=27\pi G^2_NM^2-\beta\frac{26624\pi^4}{729G_NM^4}+\mathcal{O}\left(\beta^2\right)~.
\end{equation}
Finally, using \eqref{eq:temperature} for the temperature, the mass loss rate is
\begin{equation}\label{eq:mass_loss}
    \frac{dM}{dt}=\xi \rho \left(\frac{27}{4096\pi^3G^2_NM^2}+\beta\frac{18851}{93312G^5_NM^8}\right)+\mathcal{O}\left(\beta^2\right).
\end{equation}
Integrating \eqref{eq:mass_loss} we obtain the lifetime of the black hole as
\begin{equation} \label{eq:lifetime}
    t=\frac{4096\pi^3 G^2_NM^3}{81\xi\rho}-\sqrt{\beta}\frac{8192\pi^{11/2}\sqrt{\frac{37702}{3}}\sqrt{G_N}}{6561\xi\rho} 
    +\beta\frac{2470838272\pi^6}{1594323G_NM^3\xi\rho}+\mathcal{O}\left(\beta^{3/2}\right).
\end{equation}
\section{Conclusion}\label{sec:sec4}
Our main new results are given by (i) the metric solution \eqref{eq:metric} and \eqref{eq:solution_a(r)} in the first order
with respect to the coupling constant $\beta$, (ii) the displaced black hole horizon \eqref{eq:horizon}, (iii) the deformed
Hawking temperature \eqref{eq:temperature}, (iv) the deformed entropy \eqref{eq:entropy}, (v) the non-vanishing pressure \eqref{eq:pressure}, and (vi) the lifetime \eqref{eq:lifetime}, all derived for the Schwarzschild black holes in
the SBR gravity \eqref{eq:sbr_action} that, in turn, represents the possible quantum gravity low-energy effective action
inspired by the Starobinsky inflation model and superstrings/M-theory.

In particular, it follows from  \eqref{eq:horizon} that the black hole horizon radius is decreased. It follows from 
\eqref{eq:temperature} that the Hawking temperature is increased. The black hole entropy  \eqref{eq:entropy} is increased also, whereas the black hole pressure \eqref{eq:pressure} becomes positive. As regards the black hole
lifetime \eqref{eq:lifetime}, we find two additional contributions with the opposite signs, where the negative contribution
given by the 2nd term apparently dominates but the last term may become important for black holes with small masses.
However, our approximation is expected to break down for very small black hole masses because in that case the even higher orders (in the spacetime curvature) have to be taken into account in the effective quantum gravity action.


\section*{Acknowledgements}
\noindent S.V.K. was supported by Tokyo Metropolitan University, the Japanese Society for Promotion of Science under the grant No. 22K03624,  and the Tomsk State University development program Priority-2030.

\bibliographystyle{utphys}
\bibliography{references}

\end{document}